\documentclass[10pt,a4paper]{article}
\usepackage[utf8]{inputenc}
\usepackage{amsmath}
\usepackage{amsfonts}
\usepackage{amssymb}

\usepackage{times,epsf, graphicx} 
\def\be{\begin{equation}}
\def\ee{\end{equation}}
\def\bea{\begin{eqnarray}}
\def\eea{\end{eqnarray}}
\def\rx {\rho_{\text{eff}}}

\def\orc{\Omega_{r_c}}
\def\om{\Omega_{\text{m}}}

\def\oq{\Omega_{\text{q}}}

\def\wq {w_{\text{q}}}

\author{S.Natarajan \\ Sree Sevugan Annamalai College \\Devakottai\\ R.Chandramohan ,R.Swaminathan\\Sri Vidyaagiri College \\ Puduvayal \\natrajangravity@gmail.com}
\title{Conformal cyclic evolution of phantom energy dominated universe}
\date{}

\begin{document}
\maketitle

\begin{abstract}
From the Wheeler-DeWitt solutions, the scale factor of the initial universe is discussed. In this study, scale factors from Wheeler-DeWitt solutions, loop quantum gravity, and phantom energy dominated stages are compared. Certain modifications have been attempted in scale factor and quantum potentials driven by canonical quantum gravity approaches. Their results are discussed in this work. Despite an increment of phantom energy density, avoidance of Big Rip is reported. Scale factors predicted from various models is discussed in this work. The relationship between scale factors and the smooth continuation of Aeon is discussed by the application of conformal cyclic cosmology. Quantum potentials for various models are correlated and a correction parameter is included in the cosmological constant. Phantom energy dominated, final stage non-singular evolution of the universe is reported. Eternal increment of phantom energy density without interacting with dark matter is found for the consequence of the evolution of the future universe. Also, the non-interacting solutions of phantom energy and dark matter are explained. As the evolution continues even after the final singularity is approached, the validity of conformal cyclic cosmology is predicted. Non zero values for the scale factor for the set of eigenvalues are presented. Results are compared with supersymmetric classical cosmology. The non-interacting solutions are compared with SiBI solutions.\\
\textbf{Keywords}\\ Phantom energy , Wheeler-DeWitt theory, Scale factor quantization, Loop quantum cosmology , Cosmological constant, Non interacting phantom cosmology \\
 \textbf{PACS} \\ 98.80.Qc, 02.40.Xx, 98.80.k,  98.80.Bp

\end{abstract}
\section{Introduction}

The scale factor determines the evolution of the universe. At the initial stages, it is closest to its minimum value. At later stages, the scale factor approaches infinity, classically. Here, table 1 shows the divergence of various singularities.

\begin{table*}
\label{t1}
\caption{Divergence of various cosmological parameters  }
\begin{center}
\begin{tabular}{|c|c|}
\hline
Type  & Divergence  \\
\hline
I &  a,$\rho$,p \\ 

II  & p  \\     
 
III  & $\rho$,p  \\ 

IV  & $\dot{p}$  \\ 
\hline 
\end{tabular} 
\end{center}
\end{table*}

The classical cosmological model does not provide detailed information about the dynamics of the universe at the singularity, due to the vanishing scale factor. Particularly, the classical analysis of the singularity does not make sense. The Wheeler-DeWitt analysis and the Loop Quantum Cosmology (LQC) provides detailed information about the Hubble parameter at the initial stages of the universe.

The initial singularity is a mysterious problem The classical analysis did not resolve the initial singularity; quantum treatment is required. With the help of background-independent resolution, the Loop Quantum Gravity represents a tool to understand the quantum nature of the Big Bang.  Time and critical density dependence of the scale factor are determined by loop quantum cosmology. Similarly, pre-Big Bang scenario is characterized by the conformal cyclic cosmology. This confirms the cyclic evolution of the universe. In such resolutions, the conservation of information is required to be understood. In recent days, Penrose expressed his ideas about the confirmation of the existence of Hawking points from previous Aeon, through the CMB radiation \cite{cq1}.

 The classical singularity is a mathematical point, in which the cosmological parameters such as energy conditions,
 pressure, and curvature face divergences \cite{vg1}. Every classical cosmological scenario avoids the initial singularity. The singularity can be analyzed solely through a quantum mechanical perspective.  The quantum mechanical solutions provide the understanding of initial conditions of the universe. Dynamics of the universe are approached with many theoretical models such as Big Bang \cite{unb1},  Ekpyrotic \cite{unm1}, cyclic\cite{unm3},  LCDM \cite{lcd1} and many more.  The initial conditions of the universe resemble quantum mechanical vacuum fluctuations\cite{qv1}. The Wheeler-DeWitt model was proposed for the quantization of the gravity \cite{uvac1.5}. It is applied on the quantum analysis of initial stages of the universe. 
 
 From such solutions, the universe is said to be created spontaneously from quantum vacuum \cite{uvac1}. The quantum universe that pops out of a vacuum does not require any Big Bang-like singularities \cite{vc39}. Similarly, such a scenario requires no initial boundary conditions. In contrast to other models, LQG makes consistent results over renormalization of quantum parameters for quantizing  gravity, to avoid infrared ultraviolet divergences \cite{sin1.30}. Loop Quantum Cosmology resolves the classical singularity and examines the universe quantum mechanically \cite{sin1}. In LQC, the quantum nature of the Big Bang is analysed \cite{sin1.18}. The loop quantum model suggests that at singularity, the quantum universe bounces back as the density approaches the critical level. Hence, the Big Bang is replaced with a Big Bounce in a loop quantum cosmological scenario \cite{lc9}. The formation of a quantum mechanical universe from the quantum vacuum is discussed here with the help of the Wheeler-DeWitt theorem. With scale factor quantization, the behaviour of the quantum cosmological constant is also discussed. A time-varying cosmological parameter is proposed to have a consistent value for the cosmological constant, which is currently a need of the hour. The scale factor solutions are discussed for the $k=0$ model. 
 
  To construct a local supersymmetric quantum cosmological model, an alternative procedure is
presented in \cite{lov5}.  A superfield formulation is introduced and applied to
the Friedmann-Robertson-Walker (FRW) model. Scalar field cosmologies with perfect fluid in a Robertson-Walker metric are discussed in \cite{lov6}. Asymptotic solutions for the final Friedmann stage with simple potentials are found in the same work. It also predicts that the perfect fluid and curvature may affect the evolution of the universe.
Scalar phantom energy as a cosmological dynamical system is reported in \cite{lov7}. Three characteristic solutions can be identified for the canonical formalism. Effect of local supersymmetry on cosmology is discussed in \cite{lov8}. The authors discuss the case of a supersymmetric FRW model in a flat space in the superfield formulation.

The interactions of phantom energy and matter fluid are reported in \cite{phans1}.  Avoidance of cosmic doomsday in front of phantom energy is reported in the same. What will happen for the universe, soon after such singularity is approached? Will Type I singularity stop the evolution of the universe?  Will it be the endpoint of all universes? Answers for these questions are discussed in this work.  Evolution of the universe is explained by the various cosmological models.

Initially, this work starts with a basic introduction on conformal cyclic cosmology. In the next section, an introduction to phantom energy and dark energy is provided. In the next section,  an overview for loop quantum cosmology is discussed. We discuss quantization procedures. In the next section, we analyze the solutions for Wheeler-DeWitt equations.A comparison between Wheeler-DeWitt cosmology and classical supersymmetric cosmology is drawn. In later sections, we provide solutions for non-interacting dark energy which leads to the continued evolution of the universe without approaching the final singularity.  

\section{Conformal Cyclic Cosmology}

The Conformal Cyclic Cosmology (CCC) provides alternative explanations for existing cosmological models \cite{cq1.1},\cite{cq1.2},\cite{cq1.3}. The CCC predicts the evolution of the universe as cycles or Aeons.  Each Aeon starts from a Big Bang and ends up with a Big Crunch. The CCC follows conformal structure over the metric structure. The CCC approaches the initial stages of the universe as a lower entropic state, which was higher for the remote part of the previous Aeon \cite{cq1.3}. The universe without inflation is predicted by conformal cyclic cosmology. The CCC explains the imbalance between the thermal nature of radiation and matter in the earlier phases of the universe. The suppression of gravitational degrees of freedom is a consequence of remote past Aeon which has low gravitational entropy. Thus the universe is proposed as the conformal evolution over the Aeon. It is possible to cross over the Aeons, due to the conformal invariance of physics of massive particles which survived from past Aeon. 
 The materials of galactic clusters remain in the form of Hawking radiation. Due to the evaporation, massive particles can be available as photons for the future Aeons, after the survival from Big Bang singularity.  The cosmological constant $\Lambda$ is suggested as invariant in all Aeons. In CCC the inflation is said to be absent.  The $\Lambda$  overtakes the inflation that provides exponential expansion. The CCC is consistent with LCDM \cite{lco1}, without implementing inflation. The CCC does not violate the second law of thermodynamics \cite{cq7}. The conformal cyclic cosmological model topologically analyses the past and future singularities as surfaces with smooth boundaries. Hence, future and past boundaries are equated with the same kind of topology. Relationship between past and future singularities mainly exist in the scale factor. Though the singularities come under various types, the initial Big Bang and future Big Rip are compared in table \ref{t2}. This comparison is of interest for the Big Rip is considered as the end of the evolution of the universe. So the scale factors from Big Bang and Big Rip can be mapped conformally with each other. Comparisons of scale factor from loop quantum cosmology and classical super symmetrical solutions are discussed in this work. 

Information on the existence of Hawking radiation from previous Aeon can be obtained from Hawking points. Such Hawking radiation obtained from Hawking points might be the remnant of the super-massive blackhole of the previous Aeon. Such existence can be proven by mathematical and experimental solutions such as cosmic microwave background \cite{cq1}. The conformal cyclic cosmology is theorized with the Aeons. The universe evolves as Aeons. 
The evolution is marked by an initial singularity, which undergoes an exponential expansion until it reaches the critical density and then, ends up as a singularity again. Before proceeding into such mappings, some introduction about fundamental topics are required. 
\section{Dark energy and Phantom energy}
The dark energy is a mysterious form of the energy content of the universe.    The  equation of state $\omega$ determines the acceleration of the universe. From Einstein's field equations, the cosmological constant is proposed for the consequences for the accelerated expansion of the universe \cite{efen1}. In general, the cosmological constant value is smaller than the expected value from quantum gravity. Though technical conflicts exist between theory and experiment, the cosmological constant is included in Einstein's field equations.

 The phantom energy has the values of equation of state parameter  $\omega <-1$. The phantom energy violates the null energy conditions. The phantom energy is proposed as a candidate for the sustainability of traversable wormholes \cite{phan01}. Wormholes metric with the dominance of phantom energy is studied in \cite{phan02}. Symmetric distribution of phantom energy in a static wormhole is also reported.  The phantom energy has super-luminal properties. Such properties are similar to those predicted by supergravity or higher derivative gravitational theories \cite{27.deu2} \cite{28.deu2} \cite{29.deu2}  .

In the evolutionary phase of the universe, the singularities can be classified by the diverging parameters. Big Rip singularity appears in a finite time. The finite-time singularity will occur with weak conditions of $\rho>0$  and $\rho+3P>0$ in an expanding universe. 

\begin{table*}
\label{t2}
\caption{Types of singularities}
\begin{center}

\begin{tabular}{|c|c|}
\hline 
Type & Singularity \\ 
\hline 

Type 0 & Big Crunch or Big Bang singularity \\
Type I & Big Rip singularity \\
Type II & Sudden future singularity \\
Type III & Future scale factor singularity \\
Type IV & Big separation singularity \\
Type V & $\omega$ singularity, little rip pseudo singularity \\

\hline
\end{tabular} 
\end{center}
\end{table*}

 A Type I singularity has a scale factor divergence along with the associated to energy density and pressure. Type II singularities are referred to as sudden feature singularities. Divergence of pressure with finite scale factor and energy density occurs. Type III is referred to as big freeze singularity. In Type III singularities, energy density and pressure diverges with a finite scale factor. Type IV singularity is referred to as big separation singularity, where energy density pressure and scale factor  remain finite, but the time derivative of pressure or energy density diverge. The universe can be extended after it approaches singularities of either Type II or IV, because they are relatively weak. The strong Big Rip singularity is resolved here using loop quantum cosmology. Among these singularities,  Type zero is experienced by the universe at the earlier stages and Type I  is encountered at later times.

 Various cosmological singularities are reported in \cite{lov12}. The scale factor has the following characterization,

\begin{equation}
a(t) = c_0|t-t_0|^{\eta_0}+c_1|t-t_0|^{\eta_1}+ ...,
\end{equation}
where 
$\eta_0<\eta_1<... c_0>0$.

The geodesics are parametrized as 
\begin{equation}
t' = \sqrt{\delta + \frac{P^2}{a^2(t)}}
\end{equation}

and 
\begin{equation}
r'= \pm \frac{P}{a^2(t) f(r)}
\end{equation}

with
\begin{equation}
f^2(r) = \frac{1}{1-k r^2}
\end{equation}

It has been used for simplicity a constant geodesic motion $P$ and $\delta =0$ and $1$ for null and time like geodesics, respectively. We have, for full geodesics,

\begin{eqnarray}
a(t) t' = P = \int^t_{t_0} a(t) t'\\
P= P(\tau - \tau_0),
\end{eqnarray}

If $\eta_0>0$, the scale factor vanishes at $t_0$. Hence, there will be either a Big Bang or Big Crunch. If $\eta_0=0$, then the scale factor will be finite at $t_0$. A sudden future singularity will appear in the evolution of the universe. If $\eta_0<0$, then the universe will face Big Rip at $t_0$.  If the cosmological models exist with $\eta_0\leq -1$, then the null geodesics will avoid the Big Rip singularity. In our solutions, $\eta_0 \to (-\infty,0)$, $\eta_1 \to (\eta_0,\infty)$, $k= 0, \pm 1$,and  $c_0 \to (0,\infty) $. Various cosmological singularities \cite{luc1}, \cite{luc2} are reported in table \ref{tab1}.

\begin{table*}
\label{tab1}
\caption{Strength of singularities is determeined by the various parameters}
\begin{center}

\begin{tabular}{|c|c|c|c|c|c|}
\hline 
${\eta_{0}}$ & ${\eta_{1}}$ & ${k}$ & $c_{0}$ &\textbf{Tipler} &
        \textbf{Kr\'olak}\\ 
\hline 

        $(-\infty,0)$ & $(\eta_{0},\infty)$ &  $0,\pm 1$  & $(0,\infty)$& 
        Strong & Strong  \\
        $0$ & $(0,1)$ &  $0,\pm 1$  & $(0,\infty)$ & Weak & Strong  \\
    0     & $[1,\infty)$ & $0,\pm 1$  &$(0,\infty)$& Weak & Weak
    \\
        $(0,1)$ & $(\eta_{0},\infty)$ &  $0,\pm 1$  & 
        $(0,\infty)$  & Strong & Strong  \\
        $1$ & $(1,\infty)$ & $0,1$  &  $(0,\infty)$& Strong & Strong  \\
        1 & $(1,\infty)$ & $-1$ &$(0,1)\cup(1,\infty)$ & Strong & Strong  \\
        1 & $(1,3)$ & $-1$ & 1 & Weak & Strong  \\
        1 & $[3,\infty)$ & $-1$ & 1 &Weak & Weak  \\
        $(1,\infty)$ & $(\eta_{0},\infty)$ & $0,\pm 1$  &$(0,\infty)$&
    Strong & Strong \\

\hline
\end{tabular} 
\end{center}
\end{table*}

$\omega=const$ is not the only option to obtain an accelerated cosmic expansion. There are dark energy parameterizations that can cross the phantom divided-line with success and also reproduce $\omega=-1$.

Six bidimensional dark energy parameterizations are studied and tested with available SNe Ia and BAO data \cite{lov13}. Obtained results are in favour of the LCDM model.  Various parametrizations are reported in the same reference.
The Friedmann-Raychaudhuri equation is
\begin{equation}
E(z)^2= \left(\frac{H(z)}{H_0} \right)^2= 8 \pi G (\rho_m+\rho_{DE})[\Omega_{0m}(1+z)^3+ \Omega_{0 DE} f(z)]
\end{equation}

\begin{equation}
\frac{\ddot{a}}{a}= \frac{H^2}{2}[\Omega_m+\Omega_{DE}(1+3\Omega)]
\end{equation}

where $H( z )$ is the Hubble parameter, $G$ the gravitational constant, and the subindex $0$ indicates
the present-day values for the Hubble parameter and matter densities.
For dark energy, 
\begin{equation}
\rho_{DE}(z)= \rho_{0(DE)}f(z),
\end{equation}

with 

\begin{equation}
f(z)= exp \left[3\int^{z}_{0}{\frac{1+w(\tilde{z})}{1+\tilde{z}}}d\tilde{z}\right]
\end{equation}

For quiessence models, $w = const.$

Solution of $f(z)$ is therefore,
\begin{equation}
f(z)=(1+z)^{3(1+w)}
\end{equation}
For cosmological constant, $w=-1$ and $f =1$.

\begin{table*}
\label{tab2}
\caption{Dark energy parametrizations with best fits and $\sigma-$distances values using SNe Ia JLA data. 
Referred from \cite{lov13}}
\begin{center}

\begin{tabular}{|c|c|}
\hline 
 { Model}        &{ Parametrizations }  \\  
\hline 
LCDM & $ H^2(z)=H_0^2 [\Omega_{m}(1+z)^3 + (1- \Omega_{m})]  $        \\
Linear & $H^2 (z)=H_0^2 [\Omega_{m} (1+z)^3 + (1-\Omega_{m})(1+z)^{3(1+w_0+w_1)}e^{-3 w_1 z}]$\\ 
CPL & $H^2 (z)=H_0^2 [\Omega_{m} (1+z)^3 + (1-\Omega_{m})
(1+z)^{3(1+w_0+w_1)}$
 $\times e^{-3 w_1 z\over{1+z}}] $ \\
 BA & $H^2 (z)=H_0^2 [\Omega_m (1+z)^{3}+(1-\Omega_m) (1+z)^{3(1+w_0)}$
 $\times (1+z^2)^{3w_1 /2} $ 
\\
 LC & $H^2 (z)=H_0^2  [\Omega_m (1+z)^3 + (1-\Omega_m) (1+z)^{3(1-2 w_0 +3 w_{0.5})}$  $ \times e^{\left[\frac{9(w_0 -w_{0.5})z}{1+z}\right]}] $ \\
JBP & $H^2 (z)=H_0^2 [\Omega_{m} (1+z)^3 + (1-\Omega_{m})
(1+z)^{3(1+w_0)}$  $\times e^{3 w_1 z^2\over{2(1+z)^2}}] $ \\
 WP & $H^2 (z)=H_0^2 \left\{ \Omega_m (1+z)^{3}+(1-\Omega_m)(1+z)^{3\left[1+\frac{w_0}{1+w_1 \ln{(1+z)}}\right]}\right\}$  
\\

\hline
\end{tabular} 
\end{center}
\end{table*}

In addition to the simplest models in which the universe contains only cold dark matter and a cosmological constant, class of braneworld models can lead to a phantom-like acceleration of the late universe \cite{lov14}. This model does not require  any phantom matter. The quintessence leads to a crossing of the phantom divide-line $w = 1$. This model avoids the future Big Rip by decreasing the Hubble parameter.

 For $\wq>-1$, a smooth crossing of
the phantom divide occurs at a redshift $z_{\text{c}}$ that
depends on the values of the free parameters $\om$, $\oq$ and
$\wq$.  $z_{\text{c}}$ is obtained as
 \be
(1+z_{\text{c}})^{3\wq} E(z_{\text{c}}) =
\frac{\om\sqrt{\orc}}{(1+\wq)\oq}\,.\label{crossz}
 \ee
For LDGP (Cosmological constant included Dvali-Gabadaze-Porrati) model $\wq=-1$, the crossing occurs at $a=\infty$.

At a redshift $z_*>z_{\text{c}}$, we have $\rx(z_*)=0$. Hence, the  phantom general relativistic picture of QDGP (Quintessence included Dvali-Gabadaze-Porrati) diverges.
 \be
z_*=\left(\frac{4 \orc\om}{\oq^2}\right)^{1/3(21+\wq)}-1\,.
 \ee

The future Big Rip can be avoided due to the parameters $H,\dot H \to 0$ as $a\to
\infty$. This asymptotic behavior reflects the fact that the
phantom effects might have been dumped. The total equation of state parameter is defined
by
\begin{equation}
1+\omega_{tot}(z) = \frac{\Omega_m(1+z)^3+(1+\omega_q)\Omega_q(1+z)^{3(1+\omega_q)}}{E(z)[\sqrt{\Omega_{r_c}}+E(z)]}
\end{equation}
This shows that $\omega_{\text{eff}}(z)\geq -1$.

For the quantization of the singularity, the basic understanding of quantum geometry is required. The quantum geometric analysis, done via loop quantum geometry is discussed in the next session.

\section{Loop quantum cosmology - a brief analysis} 

The canonical quantum gravitational formalism is based on the quantization of the metric. It attempts to quantize the phase space as a Hilbert space. In canonical formalism, the phase space variables are replaced by operators. But the formalism faces some constraints. There are  Hamiltonian, diffeomorphism and Gauss constraints. To solve these constraints, LQG is proposed. The model has numerous successes in resolving the initial singularity. The model provides background free solutions for the canonical gravity approaches. The Loop Quantum Gravity explains the discreet nature of the spacetime. On cosmological scales, Loop Quantum Cosmology is proposed for the unperturbed evolution of the universe. 

 Loop Quantum Cosmology is based on the canonical quantum gravitational formalism. The Loop Quantum Gravity does not require renormalization; this makes LQG stand special over other quantization approaches. From canonical quantization, one can expect non zero discrete values for geometric quantities as quantum observables. Hence, the loop quantization approach provides non zero values for area and volume. Area and volume in lLQG are formulated as operators. 
 
 \begin{eqnarray}
 A \to \hat{A}, \\
 V \to \hat{V}
 \end{eqnarray} 
 
 The constraints reduce the possibilities of quantization. Setting constraints as $c=0$,  the solution of the quantum evolution will be
 \begin{equation}
 \hat{c}_1 \psi = 0
 \end{equation}

In general, every cosmological scenario faces singularities during the initial stages of evolution. The initial conditions of the universe possess strong singularity. The initial Big Bang singularity bears high curvatures and energy density divergences. To resolve this singularity conditions, LQC is equipped with the application of LQG theory. Loop Quantum Gravity attempts to derive non-perturbative and background independent quantization of general relativity. In LQC there is a straightforward link between full theory and the cosmological models, which is in contrast to other cosmological approaches.  The full theory of quantum gravity is required to be constructed. The Loop Quantum Cosmology is based on symmetrical reduction. But this methodology faces mathematical problems in full theory. So, current research is happening on symmetrical, non-symmetrical models and their relationships.

The Robertson-Walker metric for a flat $(k=0)$ homogeneous isotropic universe in LQC is  

\begin{equation} 
ds^2=-dt^2-a^2(t)(dr^2+r^2(d \theta^2 + \sin^2 \theta d \phi^2))
\end{equation}

with $a(t)$ scale factor and $t$ is the proper time. The effective Hamiltonian in LQC is 

\begin{equation} 
\mathcal{H}_{\mathrm{eff}} = \frac{-3}{8 \pi \gamma^2 G} \frac{\sin^2 (\lambda \beta)}{\lambda^2} V
\end{equation} 

The details of quantum dynamics are provided by the effective Hamiltonian, and $\mathcal{H}_{\mathrm{matt}}$ provides the matter Hamiltonian. Some new quantum variables, $\beta$ and $V$, can be introduced in the quantum regime. The conjugate variables $\beta$ and $V$ satisfy the commutation relation  

\begin{equation} 
\{\beta, V \} = 4 \pi G \gamma,
\end{equation} 

Where $V=a^3$ and $\gamma=0.2375$ is the Barbaro-Immirizi parameter \cite{lc9.93}. The phase space variable from classical dynamics is

\begin{equation} 
\beta = \gamma \frac{\dot{a}}{a}
\end{equation} 

The parameter $\lambda$ determines the minimum eigenvalue parameter of LQG and the discreteness of quantum geometry \cite{sin1.39}. The parameter is denoted as  

\begin{equation} 
\lambda = 2(\sqrt{3} \pi \gamma)^{\frac{1}{2}} l_{pl}
\end{equation}

If the Hamiltonian constraint vanishes, i.e. $\mathcal{H}_{\mathrm{eff}}=0$, it leads to

\begin{equation} 
\frac{\sin^2 (\lambda \beta)}{\lambda^2} = \frac{8 \pi \gamma^2 G} {3} \rho
\end{equation}

with

\begin{equation}
 \rho = \frac{\mathcal{H}_{\mathrm{eff}}}{V},
\end{equation}

the energy density. From the Hamiltonian equations

\begin{equation} 
\dot{V}= V \mathcal{H}_{\mathrm{eff}} \\ = - 4 \pi G \gamma \frac {\partial}{\partial \beta} \mathcal{H}_{\mathrm{eff}} \\= \frac{3}{\gamma} \frac{\sin (\lambda \beta)}{\lambda} \cos (\lambda \beta) V
\end{equation} 

This equation can call the modified Friedmann equation as  

\begin{equation}
H^2 = \frac{\dot{V}^2}{9V^2} = \frac{8 \pi G}{3} \rho(1-\frac{\rho}{\rho_{crit}}),
\end{equation}

where the critical density is inferred as
\begin{equation}\label{a1}
\rho_{crit}= \frac{3}{8 \pi G \gamma^2 \lambda^2} \sim 0.41 \rho_{pl}
\end{equation}

 Similarly, the Raychaudhuri equation is also modified with the help of $\dot{\beta}=\beta \mathcal{H}_{\mathrm{eff}} $
\begin{equation}
\frac{\ddot{a}}{a}= \frac{-4 \pi G}{3} \rho(1-\frac{\rho}{\rho_{crit}}) - 4 \pi \rho G (1- 2\frac{\rho}{\rho_{crit}})
\end{equation}

 which holds the conservation law 
\begin{equation}
\dot{\rho}=3H(\rho+p).
\end{equation}

where the pressure is, 

\begin{equation}
p=\frac{-\partial{H_{matt}}}{\partial{V}}
\end{equation} 

 Hubble parameter in gravitational Hamiltonian is replaced by the holonomies. They are non linear functions of $a$ and $\dot{a}$. The spacetime behaves with discreteness, near the Planckian scale. Quantum replacement for classical size $a^3v_0$ is written as $n v$ \cite{lc8}. 
\begin{equation}
n(t)v(t) = a(t)^3 V_0,
\end{equation}

where $v$ is the mean size units in 3D space and $n$ the number of sections in the region. Implementation of discreteness of quantum geometry results in the dynamics. It depends on the patch size and independent of the number of patches $n$. Holonomies are represented as

\begin{equation}
\tilde{C}= \gamma \dot{a}
\end{equation}

These modified equations express the Ricci flow on FRW background. We may discuss about curvatures. Invariant form of Ricci curvature will be
\begin{equation}\label{ricci1}
R= 6(H^2+\frac{\ddot{a}}{a}) \\= 8 \pi G \rho (1-3\omega+2 \frac{\rho}{\rho_{crit}}(1+3\omega))
\end{equation}

From the equation (\ref{ricci1}), it can be observed that the curvature scalar approaches negative values for the chosen parameters such as $\rho =\rho_{crit}$ and $\omega<-1$.  Hence Anti-Desitter kind of future universe may appear. Also, the conformal Aeon will face regulated future singularity as anti-de Sitter (AdS) like singularity. 

Similarly, the Ricci components can be written in terms of lapse function $N$ and scale factor as defined in \cite{lov1},

\begin{equation}
R = 6 \left(\frac{\ddot{a}}{N^2 \ a} + \frac{\dot{a}}{N^2 \ a^2}+ \frac{k}{a^2} - \frac{\dot{a}}{a}\frac{\dot{N}}{N^3} \right)
\end{equation}

The Friedmann equation can be obtained in terms of Ashtekar variables as 

\begin{equation}
H = - \frac{3}{8 \pi G} (\gamma^2 (c -\Gamma)^2 +\Gamma^2) \sqrt{|p|} \\ + H_{matter}(p) = 0,
\end{equation}

where 
\begin{equation}
\Gamma = V_0^{\frac{1}{3}} \tilde{\Gamma}
\end{equation}

\begin{equation}
\tilde{\Gamma} = \tilde{c} - \gamma \dot{\tilde{a}}
\end{equation}

and 
\begin{equation}
c = V_0^{\frac{1}{3}} \tilde{c}
\end{equation}

In general, the AdS universe expands eternally. But there exists a maximum cut-off by loop quantum cosmological solutions. Those solutions predict that when the energy density reaches $\rho_{crit}$ then the universe will nucleate future Aeon. 

With
\begin{equation}
\omega=\frac{p}{\rho},
\end{equation}

as $\lambda \to 0$ which  results in $ G \hbar \to 0$, then the equations will behave classicaly as
\begin{equation}
H^2= \frac{8 \pi G}{3} \rho
\end{equation}
\begin{equation}
\frac{\ddot{a}}{a}= \frac{-4 \pi G}{3}(\rho+3p)
\end{equation}
\begin{equation}
R= 8 \pi G (\rho+3p)
\end{equation}
\cite{deu5.12}\cite{deu5.13}

 In LQC, the scalar field is considered as an internal clock. The Hubble distance is defined as
\begin{equation}
H^{-1}=\frac{a}{\dot{a}}
\end{equation}
In LQG formalism, the canonical part satisfies
\begin{equation}
\{H, v_0a^3\}= 4 \pi G
\end{equation}

 The scale factor is quantized with  Ashtekar variables \cite{lc9} as follows,
 \begin{equation}\label{sclq1}
 A^i_a= c V_0^{-\frac{1}{3}} \mathring{\omega}^i_a
 \end{equation}
 Its conjugate momenta is represented as
 \begin{equation}\label{sclq2}
 E^a_i = p V_0^{-\frac{2}{3}} \sqrt{\mathring{q}} \mathring{e}^a_i
 \end{equation}
 
 with 
 \begin{equation}
 |p|= V_0^{\frac{2}{3}} a^2
 \end{equation}
 \begin{equation}
 c= \gamma V_0^{\frac{1}{3}} \frac{\dot{a}}{N}
 \end{equation}
 which satisfies 
 \begin{equation}
 \{c,p\}= \frac{8 \pi G \gamma}{3 V_0}
 \end{equation}
 Here, $\mathring{\omega}^i_a$ and $\mathring{e}^a_i$ are fiducial triads and $q_{ab}$ is fiducial metric. These equations reveal the relationships between triads and scale factors.
 \begin{equation}
 b= \frac{c}{|p|^{\frac{1}{2}}}
 \end{equation}
 \begin{equation}
 v= sgn(p) \frac{|p|^{\frac{3}{2}}}{2 \pi G}
 \end{equation}
 
 The holonomy $ e^{\left(\imath \frac{l_0c}{V_0^{\frac{1}{3}}}\right)}$ becomes shift operator in $p$. The quantization is confirmed by promoting Poisson bracket into commutator.
The Poisson bracket of the volume with connection components can be represented as 

\begin{equation}
\{A_a^i, \sqrt{|det E|} d^3 x \} = 2 \pi G \epsilon^{ijk} \epsilon_{abc} \frac{E_j^B \ E_k^c}{\sqrt{|det E|}}
\end{equation} 

The metric of an isotropic slice is
 
\begin{equation}\label{tri1}
q_{ij}=a^2\ \delta_{IJ}\ =\ e^i_I\ e^i_J
\end{equation} 

Here $e^i_I$ is a co-triad. The geometry of spatial slice $\sum$ is encoded in the structural metric $q_{ab}$. The canonical momenta $K_{ab}$ is driven by extrinsic curvature. Diffeomorphisms constraints produce deformations of a spatial slice. This may be connected to the decoherence of quantized space in the singularity. In LQG the geometrical formalism is defined by the triad formalism $e_i^a$, not by the spatial metric. Summations over the indices connect the triads vector fields. The advantage of choosing the triad over metric is that the triad vectors can be rotated without changing the metric. This entails additional gauge freedom with group $SO(3)$ acting on $i$.

To understand the space of metrics or structure tensors, Ashtekar variables are introduced.

Triads can be written as denstized form. That the densitized triads conjugate extrinsic curvature coefficient.

\begin{equation}
k_i^a = K_{ab} e^b_i
\end{equation}

\begin{equation}
\{k^i_a(x),E_j^b(y)\}= 8 \pi G \delta^b_a \delta^i_j \delta(x,y)
\end{equation}

The curvature is replaced with Ashteaker connections
\begin{equation}
A^i_a = \Gamma^i_a + \gamma k^i_a 
\end{equation}

Ashteaker connections conjugate to triads will be
\begin{equation}
\{A^i_a(x),E_j^b(y)\}= 8 \pi G \delta^b_a \delta^i_j \delta(x,y)
\end{equation}

Hence, the spin connection will be
\begin{equation}
\Gamma^i_a = \epsilon^{ijk} e^b_j(\partial_{[a}e^k_e]+\frac{1}{2}e^c_ke^i_a \partial_{[c}e^i_b])
\end{equation}

The spatial geometry is obtained from densitized triads,
\begin{equation}
E^a_iE^b_i = q^{ab} det \ q
\end{equation}

An expression for the inverse scale factor is
\begin{subequations}  
\begin{eqnarray}
M_{IJ} &=& \frac{q_{IJ}}{\sqrt{\det{q}}} \\
 & =&  \frac{e^i_I\ e^i_J}{(\det e)} \\ 
 &=& \frac{1}{a}\  \delta_{IJ}. 
\end{eqnarray} 
\end{subequations} 

The inverse scale factor $M_{IJ}$ can be quantized to volume operator \cite{cq2.14}. The bounded operator will be
    
\begin{equation}\label{inve1}
\begin{split}
M_{IJ,j} = \frac{16}{\gamma^2\  l_p^2} (4 \sqrt{V_j} - \frac{1}{2}\sqrt{V_{j+\frac{1}{2}}}-\frac{1}{2}\sqrt{V_{j-\frac{1}{2}}})^2 + \\ \delta_{IJ}(\sqrt{V_{j+\frac{1}{2}}}-\sqrt{V_{j-\frac{1}{2}}})^2
\end{split}
\end{equation}     
     
     Quantization of the inverse scale factor does not diverge at the singularity. Even though the volume operator diverges, the corresponding inverse scale factor does not diverge.

There are eigenstates of volume operator $\hat{V}$ with eigenvalues \cite{cq2.10},

\begin{equation}\label{vol}
V_j=(\gamma l_p^2)^{\frac{3}{2}}\sqrt{\frac{1}{27}j(j+\frac{1}{2})(j+1)}
\end{equation}
     
Those results for eigenspectrum and scale factor values are plotted in Figure 2.

 The full Hamiltonian depends upon the patch volume but not the number of patches or the total volume.
 
 \section{Wheeler-DeWitt solution for initial scale factor}

 Wheeler-DeWitt equations attempted to quantize the initial singularity. Like Einstein's field equation, Wheeler-DeWitt equation is also a field equation. The Wheeler-DeWitt approach attempts to quantize gravity by connecting General Relativity and Quantum Mechanics. The Wheeler-DeWitt equation resolves the Hamiltonian constraint using metric variables.

 Mini super-space models can be implemented to explain the emergence of the universe \cite{uvac1.7}, \cite{uvac1.8},\cite{uvac1.9}. Action of mini super-space is defined as
 \begin{equation}\label{a}
 S = \frac{1}{16 \pi G} \int R \sqrt{-g} d^4x
 \end{equation} 
 Metric of the mini super-space is defined by
 \begin{equation}\label{b}
 ds^2= \sigma^2 [N^2(t) dt^2 - a^2(t) d \Omega^2_3]
 \end{equation}
 The mini super-space is considered to be homogeneous and isotropic. Here, $d \Omega^2_3$ is metric on $3$-sphere,  $N^2(t)$ is  the lapse function. $\sigma^2$ is normalization parameter. From equations (\ref{a}) and (\ref{b}) the Lagrangian is derived as
 \begin{equation}\label{c}
 \mathcal{L}= \frac{N}{2} a \left( k- \frac{\dot{a}^2}{N^2} \right),
 \end{equation}   
 and the momentum is 
 \begin{equation}\label{d}
 p_a=\frac{-a \dot{a}}{N}
 \end{equation}
 
 The canonical form of the Lagrangian can be written as
 \begin{equation}\label{e}
 \mathcal{L}= p_a \dot{a} - N \mathcal{H},
 \end{equation}
  where
 \begin{equation}\label{f}
 \mathcal{H} = \frac{-1}{2} \left(\frac{p_a^2}{a} + ka \right)
 \end{equation} 
 The Wheeler-DeWitt  theory determines the evolution of the universe. The Hamiltonian is in the form
 \begin{equation}\label{g}
 \mathcal{H} \psi = 0
 \end{equation}  
    and 
 \begin{equation}\label{h}
 p_a^2 = -a^{-p} \frac{\partial}{\partial{a}} \left(a^p \frac{\partial}{\partial{a}}\right)
 \end{equation}   
 Then the Wheeler-DeWitt equation (WDWE) is changed as \cite{uvac1.6} \cite{uvac1.10}
 \begin{equation}\label{i}
 \left(\frac{1}{a^{p}} \frac{\partial}{\partial{a}} a^p \frac{\partial}{\partial{a}}-ka^2\right) \psi(a) = 0
 \end{equation}
 Here $K=0,+1, and -1$ for flat, closed, and open bubbles, respectively. The quantum trajectories can be obtained from quantum field theory and non-relativistic perspectives \cite{uvac1.7}\cite{uvac1.14}. It is represented as
 \begin{equation}\label{j}
 \frac{\partial{L}}{\partial{\dot{a}}} = a\dot{a} =\frac{\partial{S}}{\partial{a}},
 \end{equation}
 
 \begin{equation}\label{k}
 \dot{a} = \frac{-1}{a} \frac{\partial{S}}{\partial{a}}.
 \end{equation}

  The inflation occurs for the selected values of $p=-2$ or $4$. The quantum vacuum experiences exponential expansion which is triggered by quantum potential \cite{uvac1}. The expansion is analyzed for a flat, i.e. $k=0$ bubble. The analytic solution for equation (\ref{i}) is,
 \begin{equation} \label{dam1}
 \psi(a)= i b_1 \frac{a^{1-p}}{1-p} - b_2,
 \end{equation} 
  
 where $b_1 and b_2$ are arbitrary constants.

 The scale factor can be determined as 
 \begin{equation}\label{m}
     a(t)=
     \begin{cases}
       (\frac{b_1}{b_2}(3-|1-p|)(t+t_0))^{\frac{1}{(3-|1-p|)}}, & \ |1-p| \neq 0,3  \\
       e^{b_1 \frac{(t+t_0)}{b_2}}, &\ |1-p|=3
     \end{cases}
 \end{equation}

 As discussed earlier for $p=-2$ or $4$, then $\frac{b_1}{b_2}>0$. Quantum potential corresponding to the small scale factor is
 
 \begin{equation}\label{n}
 Q(a \to 0) = - \frac{b_1}{b_2}\frac{1}{a^3} 
 \end{equation}
    
  Hence, the classical potential $V(a)$ cancels out. The effect of quantum potential on vacuum bubbles resembles to the scalar filed potential \cite{uvac1.15} or cosmological constant \cite{uvac1.16}. The effective cosmological constant for a $k=0$ bubble is in the order of
 \begin{equation}\label{o}
 \Lambda \sim 3 \left(\frac{b_1}{b_2}\right)
 \end{equation}

 The universe will expand rapidly for a scale factor $a<<1$ and it will stop its expansion for  $a>>1$. Quantum potential plays the role of cosmological constant, which kicks off the exponential expansion. 
 \subsection{Comparison of results to supersymmetric classical cosmology}
 
 Results obtained for the function  $\psi$ are compared to the equation \ref{dam1} and solution from reference \cite{lov2}. Here, $\psi$ has obtained as the form of WKB solution.
 \begin{equation}\label{dam2}
  \psi = e^{(S_{\varphi}+S_{a})}
  \end{equation} 
  
 Comparing equation \ref{dam1} and \ref{dam2} leads to the solution of the scale factor. For supersymmetric cosmology, the solution obtained is 
 
\begin{equation}
 \dot{\varphi} (t) = \dot{\varphi}_0 \left(\frac{a_0}{a}\right)^3,
\end{equation}
 
 \begin{equation}\label{dam3}
 a^3(t) = a_0^3 + 3 \left( \frac{\kappa^2 \dot{\varphi_0^2 a_0^2}}{6} + \frac{\kappa^4}{32}\right)^{\frac{1}{2}}(t-t_0)
 \end{equation}
 Here, $\kappa^2 = 8 \pi G$
 
 From equation \ref{m}, and keeping $p=-2$, and $\frac{b_1}{b_2}> 0$ hence, 
 \begin{equation}\label{dam4}
 a(t)= e^{b_1 \frac{(t+t_0)}{b_2}}
 \end{equation}
 
 Scale factors from Wheeler-DeWitt and supersymmetric cosmology can be compared. 
 Comparing equations \ref{dam3} and \ref{dam4}, 
 \begin{equation}
  a(t) =  e^{b_1 \frac{(t+t_0)}{b_2}}  =  a_0 + \left(3 \left( \sqrt{\frac{\kappa^2 \dot{\varphi_0^2 a_0^2}}{6} + \frac{\kappa^4}{32}}\right)(t-t_0)\right)^{\frac{1}{3}}
 \end{equation}
 Scale factor of the bubble universe is compared with the classical super symmetric solutions. 
 At $t=t_0$, 
 \begin{equation}
 a(t_0)= e^{b_1 \frac{(2t_0)}{b_2}}
 \end{equation}
 
 if $\frac{b_1}{b_2}=1$ then
 \begin{equation}
 a_0 = e^{2t_0}
 \end{equation}
 
 Hence, equation \ref{dam3} becomes 
 \begin{equation}
  a(t)= e^{2t_0} + \left(3 \left( \sqrt{\frac{\kappa^2 \dot{\varphi_0^2 e^{4t_0}}}{6} + \frac{\kappa^4}{32}}\right)(t-t_0)\right)^{\frac{1}{3}}
 \end{equation}
 
 This equation provides a modified scale factor for the discussed solutions above. 
 
 \section{Non interacting solutions}
   The conformal cyclic cosmology has different solutions for the evolution of the universe. The model predicts that the universe evolves as conformal cycles. Hence, the initial singularity can be modified. As from the conformal model, the initial singularity is smooth and has finite surface. Similarly, the final singularity is also smooth with finite surface. To obtain such finiteness, the initial singularity must be subject to expansion and the final singularity must cope with contraction. Instead of future Big Crunch, one may be curious about the conformal mapping between phantom dominated final stages of the universe and initial stages of the universe. The phantom dominance at final stages of the universe are characterized below.
 
 The scale factor by phantom dominated final stages with $\omega<-1$ can be obtained from the relation
 
 \begin{equation}
 a(t) = a(t_m) \left(-\omega (1+\omega)\left(\frac{t}{t_m}\right)\right)^{\frac{2}{3(1+\omega)}}
 \end{equation}
 The scale factor will blow up with time 
 \begin{equation}
 t=\frac{\omega t_m}{(1+\omega)}
 \end{equation}
 
 Friedmann equation in terms of matter fluid $\rho_m$ and phantom field $\phi$ can be written as
 \begin{equation}
 H^2 = \frac{8 \pi G}{3} (\rho_x+\rho_m)
 \end{equation}
 
 Energy density and pressure of the phantom field is obtained by the following equations,
 \begin{eqnarray}
 \rho_x = - \frac{1}{2} \dot{\phi}^2 + V(\phi) \\
 P_x = - \frac{1}{2} \dot{\phi}^2 - V(\phi) 
 \end{eqnarray}
 Here, $V(\phi) $ indicates phantom field potential.
 
 . The interaction between dark matter and dark energy can be explained with interaction term $\Gamma$,
 \begin{equation}
 \Gamma_x = \dot{\rho_x}+3H\epsilon_x\rho_x
 \end{equation} 
 \begin{equation}
 \Gamma_m = \dot{\rho_m}+3H\epsilon_m\rho_m
 \end{equation}
 with
 \begin{eqnarray}
 \epsilon_x=1+\omega_x = -\frac{\rho_x+P_x}{\rho_x}\\
 \epsilon_m=1+\omega_m = \frac{\rho_m+P_m}{\rho_m}
 \end{eqnarray}
 
 where
 $\rho_x$ is the energy density of dark energy and $\rho_m$ the energy density of dark matter.
 
 The term $\epsilon$ has values as $\epsilon_x \leq 0$ and $1\leq\epsilon_m \leq 2$. The energy density ratio between dark matter and the phantom energy field gives additional freedom for the interaction $\gamma$,
 \begin{equation}
 r = \frac{\rho_m}{\rho_x}
 \end{equation}
 
 The interaction term is redefined here based on the phantom energy considerations. The interaction can be explained as

  . 
 \begin{equation}
 \Gamma = 3HC^2 (\rho_x+\rho_m),
 \end{equation}
  where $C$ is the coupling constant \cite{phans3}. When the coupling constant is positive, the energy will be transferred from dark energy to dark matter.  When the coupling constant is negative, the dark energy will be transferred from dark matter to dark energy.
  
  A similar kind of interaction can be found in \cite{phans1}, 
  \begin{equation}
  \gamma = 3H (\epsilon_m-\epsilon_x) \frac{r \rho_x}{1+r}
  \end{equation}
  
  with
  \begin{equation}\label{anm1}
  \epsilon_m = 1+ \omega_m
  \end{equation}
  \begin{equation}\label{anm2}
  \epsilon_x = 1+ \omega_x
  \end{equation}

 The interaction parameter to be considered, While phantom and dark energy crosses the phantom divide line.

  The universe will avoid phantom dominated Big Rip due to the interaction between dark matter and the phantom energy. Energy conversion allows phantom energy to be transformed into dark matter. Then the universe is suggested to face an accelerated expansion phase.
  
  For the universe to continue its evolution by adding up phantom energy, the interaction between dark matter and phantom energy must be nullified;  this interaction depends on the coupling constant
  
  \begin{equation}
  C= r \frac{1-\epsilon_x}{(1+r)^2},
  \end{equation} 
  
  which determines the type of interaction. If the coupling constant is positive transformation of energy between phantom energy to dark matter happens. The positivity of the coupling constant is confirmed by $\epsilon_x<1$. Hence, the future increment of phantom energy density will be deduced. But CCC and LQC require the final state of the universe to be equivalent to the initial density by the topological structure, so the eternal increment of the phantom energy is necessarily required. 
  
  Hence, the coupling constant term can be modified with
  
  \begin{equation}
 NC= Nr \frac{1-\epsilon_x}{(1+r)^2}
  \end{equation}
  
  \begin{equation}
 NC=  \frac{Nr-Nr\epsilon_x}{(1+r)^2}
  \end{equation}
  Setting LHS to zero
  \begin{equation}
 0=  Nr-Nr\epsilon_x
  \end{equation}
  Rearranging,
 \begin{equation}
 0=  Nr-Nr(1+\omega_x),
 \end{equation}
 \begin{equation}
 0=  Nr-Nr+Nr\omega_x
 \end{equation}
 Then
 \begin{equation}\label{lam0.1}
 0=  Nr\omega_x
 \end{equation}
 Setting $\omega \neq 0$ and $N \neq 0$, 
 \begin{equation}
 r=0
 \end{equation}
 or
 \begin{equation}\label{noni1}
 C=0
 \end{equation}
 
 For the case $r=0$, 
 \begin{equation}\label{ete1}
 \frac{\rho_m}{\rho_x} = 0
 \end{equation}
 
 Hence, ${\rho_x} $ has to attain maximum values as compared with ${\rho_m}$, so the interaction between phantom and cold dark matter stops yielding the existence of eternally increasing phantom energy theoretically predicted from the equation (\ref{ete1}).
 
 The phantom dominated scale factor can be written as 
 \begin{equation}\label{pasc1}
 a(t)  = a(t_m)(1-\sigma+\sigma \frac{t}{t_m})^{\frac{2}{3\sigma}}
 \end{equation}
 
 with 
 \begin{equation}
 \sigma = \frac{\epsilon_x r}{1+r}
 \end{equation}
 
 For $\omega<-1$, the scale factor approaches its maximum values with non interacting solutions ($r=0$, $C=0$). Hence,
 \begin{equation}
 a(t) \to a_{max}
 \end{equation}

 The dark energy interaction can be discussed within a LQG model. Here, $\omega_x>-1$ is quintessence mode and $\omega_x<-1$ is phantom mode.
 
 Density perturbations in the universe can also be dominated by the Chaplygin gas, which has negative pressure \cite{lov3}. In addition to non-interacting solutions of quintessence and phantom, Chaplying gases can fulfill such requirements.  The Chaplygin gas can be a possible candidate for dark energy \cite{lov4}. The  Chaplygin gas has an equation of state
\begin{equation}
P = - \frac{A}{\rho},
\end{equation}

with $A$ is a positive constant.

\section{Final stages of the universe}
 
 In classical case, values for the scale factor can be obtained from \cite{deu4}
 \begin{equation}\label{scale1}
 a = a_0 exp\left[ \frac{1}{6} \frac{(2A+B\rho^{1-\alpha})\rho^{1-\alpha}}{AB(1-\alpha)}\right]
 \end{equation}

 Depending upon the choice of parameters, future singularities will appear. The maximum value for the scale factor can be obtained from equation (\ref{scale1}). In this case, no strong singularities will appear for the values of $\frac{1}{2}<\alpha<\frac{3}{4}$. For the values $\alpha = 0.8$, $A=1$ and $B=1$, the scale factor will face extreme values. 
 
 Usually the Type I singularity faces the dominance from phantom energy. The density of the phantom energy increases over time. 
 
 Equation (\ref{scale1}) can be rewritten as 
 
 \begin{equation}\label{scale2}
 a = a_0 exp\left[ \frac{1}{6} \frac{(2A+B(\rho_m+\rho_p)^{1-\alpha})(\rho_m+\rho_p)^{1-\alpha}}{AB(1-\alpha)}\right]
 \end{equation}
 
 At the very final stages, the standard model predicts that the future universe will have a very low matter density. But the phantom energy density will increase over time. This can be understood with the help of figure 1. Therefore, the energy density will approach to maximum values.
 \begin{figure*}\label{img1}
 \begin{center}
 \includegraphics[scale=0.7]{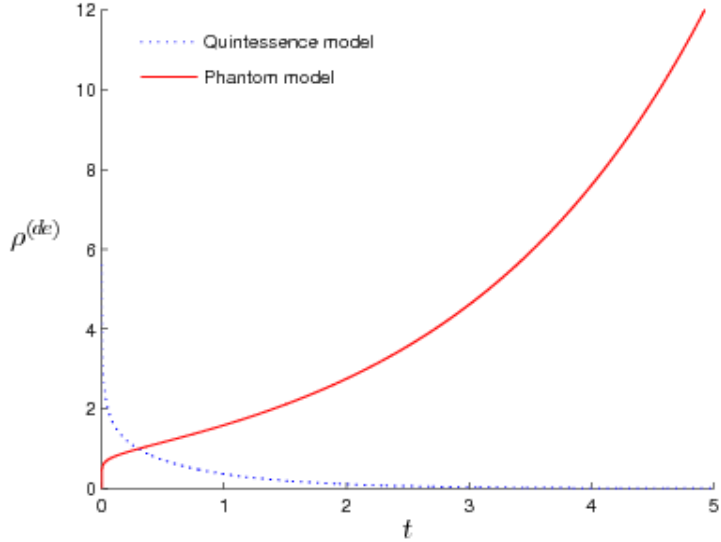}
 \caption{Increment of phantom energy over time is shown in the figure. As time increases the phantom energy keep on increases. Referred from \cite{phans2}}.
 \end{center} 
 \end{figure*}
 
 When the phantom energy density reaches values greater than the matter-energy density $\rho_p>>\rho_m$, the future universe will approach the Big Rip. For the spatially FRW universe, the Friedmann equation can be written as
 \begin{equation}
 H^2 = \frac{8 \pi G}{3} (\rho_p+ \rho_m ),
 \end{equation}
 
 where $\rho_m$ is the density of matter field and $\rho_p$ is the density of the phantom field. If the increasing phantom energy density approaches the values $\rho_p \sim \rho_{crit}$, (where $\rho_{crit} \sim 0.41 \rho_{pl}$) then the universe should bounce back as  it does in Type 0 or Big Bang singularity. As suggested from \cite{phans1}, the Big Rip singularity can be avoided. But the implementation of loop quantum modification of classical analysis is required to explain such a phenomenon. 
 
 The LQC has modified Friedmann equations \cite{lc9.26} such as

 \begin{equation}
 H^2 = \frac{8 \pi G}{3} \rho (1- \frac{\rho}{\rho_{crit}})
 \end{equation} 
 From the modified Friedmann equations, one can understand that the universe will bounce back once  the critical density is attained. The increasing phantom energy density may come close to Planck density values. This scenario is obtained by the non-interacting phantom energy model. Then, instead of blow off, the universe will bounce back at the later times of the evolution.  
 
 Smooth mapping between the equations (\ref{sclq1}), (\ref{m}) ,(\ref{pasc1}), (\ref{scale2}) and (\ref{lqm1}) is required for the conformal evolution of the universe.  At the initial stages, the universe has the scale factor as obtained from  equations (\ref{m}) and (\ref{scale2}) and at later stages it has been modified with equations  (\ref{sclq1}) and (\ref{pasc1}).
 
 \section{Phantom dominated possible state of universe}
 At vanishing or diverging scale factors, the universe undergoes a Big Bang or Big Rip singularity, accordingly. The loop quantum universe behaves as the de Sitter universe in such regimes \cite{sin1}. The universe itself appears to be a de Sitter space instead of the universe tunnelling into de Sitter space \cite{vu1}. The dynamic and geodesic equations do not have cut offs in LQC as the energy density and Hubble rate are bonded together.  In general,  LQC invokes all singularities and that holds only weak singularities and curvature singularities.  Previous attempts to resolve the Big Bang singularity, such as the Wheeler-DeWitt theory which didn't resolve the nature of singularity. The classical trajectory of a Wheeler-DeWitt solution leads to a Big Bang singularity, that requires a different theory rather than Wheeler-DeWitt's work to solve the Big Bang singularity. Loop Quantum Cosmology satisfies such requirements. It resolves all the classical singularities and elaborates the possibility of extension of space beyond the classical singularity.

  From equation \ref{a1}, it has been shown that the critical density will be in the order of $\sim 0.41 \rho_{pl}$ . The time dependent curvature scalar will be in the order of
 \begin{equation}
 R= 8 \pi G (\rho + 3p)
 \end{equation}
 
  In the classical Big Bang singularity, the scale factor, energy density, and the curvature invariants vanish. In a Type 0 singularity the null  energy condition $(\rho + p)>0$ is satisfied. But in a Type I singularity, the null energy conditions are violated. Despite the divergence in the existence of energy density and pressure, the Type III singularity has finiteness in the scale factor.  Hence, Type I singularity is resolved as Type III singularity and the universe will have the upper limit for the scale factor. As the universe approaches the upper limit of the scale factor, the energy density also approaches the maximum limit derived from LQC. Instead of completely ripping off, the universe will bounce back from the Big Rip, while the energy density approaches $\rho_{pl} \to 0.41 \rho_{pl}$. Hence, the Big Rip singularity is resolved. Here, in a similar way, the Hubble rate divergence may also be resolved.
  In classical Big Rip solutions, the Hubble rate diverges. Meanwhile, in LQC, the Hubble rate has its maximum numerical value as
 \begin{equation}
 |H_{max}|= \frac{1}{2 \gamma \lambda}
 \end{equation}

   with
 \begin{equation}
 \rho_{max} = \frac{3}{8 \pi G \gamma^2 \lambda^2}
 \end{equation}  
 and 
 \begin{equation}
 \lambda^2 = \Delta l_{pl}^2
 \end{equation}
 \begin{equation}
 \Delta = 4 \sqrt{3} \pi \gamma
 \end{equation}

 As per the classical evolution at the Big Rip, the universe will get a continuous increment of the scale factor, energy density, and pressure. Loop Quantum Cosmology has the maximum value for the energy density and Hubble rate. When the energy density approaches a value close to the critical density, the dynamical effects will be lead by the quantum effects. The acceleration parameter $\ddot{a}$ approaches to a negative value and the Hubble rate approaches zero. Instead of universe ripping apart by finite time, it will re-collapse and evolution will continue for future cycles of the universe. Curvature Independence Ricci scalar will be
 
 \begin{equation}
 R = 6 (H^2 + \frac{\ddot{a}}{a}+\frac{k}{a^2})
 \end{equation}
 Hence,
 \begin{equation}
 \begin{split}
 R= 6 \bigg(\frac{4 \pi G}{3} (\rho+3p)+\frac{k}{a^2} + \\ \frac{8 \pi G}{3}(\frac{\rho}{\rho_{crit}}+k \chi)(\rho+3p)\\+\frac{k \chi}{\gamma^2 \Delta}-\frac{2 \xi_k}{\gamma^2 \Delta}(\frac{\rho}{\rho_{crit}}+k \chi - \frac{1}{\rho})  \bigg)
 \end{split}
 \end{equation}
  As $\Delta \to 0$, Ricci scalar approaches zero. Here, $K$ is the  curvature index. The value of ? is different for $K=-1$ and $K=+1$ \cite{deu2.12},
   \begin{equation}
     \chi=
     \begin{cases}
       \sin^2 \mu (1+\gamma^2) \mu^{-2}, & \text{for}\ K=-1 \\
       -\gamma^2 \mu^2, & \text{for}\ K=+1
     \end{cases}
   \end{equation}
  
  \cite{deu2.27},  \cite{deu2.28}
  
  Here,
  \begin{equation}
  \Delta \to \mu^2 p = 4 \sqrt{3} \pi \gamma \l_{p}^2,
  \end{equation}
  
  this is the minimum eigenvalue of the area operator. Then,
   
 \begin{equation}
 f_{op} =\sin^2 \mu - \bar{\mu} \sin(\bar{\mu}) \cos (\bar{\mu}),
 \end{equation}
 revealing that it is possible that the Big Rip singularity can be resolved.

 \section{Relating quantum potential $\Lambda$ with N}
 
  The quantization is confirmed by promoting Poisson bracket into commutator relations. In LQC, the inverse volume quantization provides discrete values. The differentiation equation converts and difference equation,
  
  \begin{equation}
  \frac{d|p|^{\frac{3r}{2}}}{dp} \to \frac{(|p + \Delta p|^{\frac{3r}{2}} -|p - \Delta p|^{\frac{3r}{2}} )}{2 \Delta p}
  \end{equation}
  
  The Loop Quantum Cosmological model with FRW solutions has many salient mathematical advantages. The loop quantum cosmology replaces the Big Bang with a Big Bounce. There are many similarities and differences between Wheeler-DeWitt theory and LQC. In general, classical relativity works very well, either until the scalar curvature reaches $\sim \frac{0.15 \pi}{l_{pl}^2}$ or the matter density reaches $0.01 \rho_{pl}$. The classical evolution breaks down at the singularity. But the loop quantum evolution analyses such scenario, as an extension of previous cycles of the universe. Loop Quantum Cosmology introduces symmetry reduction formalism. The Wheeler-DeWitt theory agrees with LQC with finite accuracy. Loop Quantum Cosmology provides non-zero eigenvalues for the area gap $\Delta$ and it introduces the elementary cell $\mathcal{V}$. The dynamics of LQC is analyzed with the implementation of fiducial triads $\mathring{e}^a_i$ and cotriads $\mathring{\omega}^i_a$  that defines a flat metric $\mathring{q}^{ab}$.
  
  In LQC,  $p$ and $a$ are related to the scale factor as
  \begin{equation}
  c= \gamma \dot{a} 
  \end{equation} 
  
  \begin{equation}
  p= \sigma a^2,
  \end{equation} 
    
  where  $\sigma = \pm 1$ is the orientation factor. Loop Quantum Cosmology is an exactly solvable model. Hence, the scalar field is deployed as internal time \cite{lqsc1}. As per the LQC formalism, it has been proposed that the quantum bounce is generic. There is an upper bound for the matter density. There is a fundamental discreteness of space-time, which is derived from its loop quantum
  nature. Loop quantum cosmological analysis can be introduced by Wheeler-DeWitt solutions, for the universe to emerge from vacuum also. The scale factor has the values for $k=0$ model as from the equation (\ref{m}). For the universe to be created out of vacuum, the singularity can be treated with LQC. The matter bouncing scale factor from LQC is given by 
  
  \begin{equation}\label{lqm1} 
  a_{mb}= \left(\frac{3}{4}\rho_c t^2+1\right)^\frac{1}{3} 
  \end{equation} 
  
  Matter bouncing scale factor is proportional to the critical density, which is not included the Wheeler-DeWitt (WDW) solutions. The Hubble rate for the matter bouncing scenario is 
    
  \begin{equation}\label{lqm2} 
  H_{mb}(t)= \frac{\frac{1}{2} \rho_c t}{\frac{3}{4} \rho_c t^2 +1} 
  \end{equation}

   The matter density at the bouncing scenario is also derived from LQC formalisms,  
  
  \begin{equation}\label{lqm3} 
  \rho_{mb}(t) = \frac{\rho_c}{\frac{3}{4}t^2 +1} 
  \end{equation} 
  
  Universe bounces back with the values of energy density, obtained from equation (\ref{lqm3}). The cosmological constant is discussed as a quantum potential in WDW at equation (\ref{n}). The quantum potential is the cause for the accelerated expansion of the quantum vacuum bubbles. Though the cosmological constant is quantized with LQC calculations, it requires modifications to make it constant throughout the evolution. The equation (\ref{lqm3}) can be modified via the following way,
  
  \begin{equation}\label{lqm3.1}
  \rho_{mb}(t) = \frac{1}{\frac{3}{4}t^2 +1} +\rho_c
  \end{equation}
  
  This makes the energy density of the universe in initial and later times as equal. Both $t \to 0$ and $t \to \infty$ provide equal values in energy density. Hence, the Big Rip induced initial stages is also possible. The quantum potential from the WDW equation (\ref{n}) can be treated with detailed mathematical analysis. The equation (\ref{n}) can be modified with the help of equation (\ref{lam0.1}),
  
  \begin{equation}\label{lqm4} 
  \Lambda \sim - \frac{b_1}{b_2}\frac{ N (t)}{a^3} 
  \end{equation}

  here $N(t)$ is a time-varying parameter that keeps the phantom energy to be invariant throughout the evolution. The effective Hamiltonian is  
  
  \begin{equation}\label{lqm5} 
  \mathcal{H}_{lqc}= -3V \frac{\sin^2 \lambda \beta}{\gamma^2 \lambda^2} +V\rho 
  \end{equation}

  The Wheeler-DeWitt equation has the Hamiltonian as explained from equation (\ref{f}). The loop quantum version of the Hamiltonian is equation (\ref{lqm5}). Loop Quantum Cosmology renormalizes the cosmological constant quantum mechanically,

  \begin{equation}\label{lqa1} 
  \Lambda' = \Lambda \left(1- \frac{\Lambda}{8 \pi G \rho_{crit}}\right) 
  \end{equation} 
  
  This solution modifies the FRW equations as 
  
  \begin{equation}\label{lqa2} 
  H^2= \frac{\Lambda'}{3} 
  \end{equation}

  The resolution is free from the classical potential $V(a)$. The phantom energy, which is the function of critical density, is scrutinized with LQC formulations. Earlier works suggest that the quantum potential should be proportional to $a^4$. This introduces more errors in obtaining meaningful values of the cosmological constant. Hence, we have modified the $\Lambda$ parameter with equation (\ref{lqm4}).
  
\subsection{Eddington-inspired Born-Infeld theory of gravity solutions}

 The late tome universe will face the Big Rip at a finite time. Such a scenario is referred to as cosmic doomsday. It can be analysed via the modified theory of classical and quantum gravity \cite{lov9}. Eddington-inspired Born-Infeld (EiBI) singularity solutions also can play a vital role in analyzing cosmological singularities. EiBI model confirms the availability of auxiliary finite scale factor in Big Rip like singular stages.

The EiBI action is defined as  \cite{lov10} 

\begin{equation}
S_{EiBI} = \frac{2}{k} \int d^4x  [ \sqrt{|g_{\mu \nu} + R_{\mu \nu}(\Gamma)|} \lambda \sqrt{g} ] + S_m (g)
\end{equation}

where $k$ is a constant which is assumed to be positive. The Big Bang singularity is removed by the EiBI model. Similarly, the late time Big Rip (Little Lip, Sittle Sibling Big Rip) can also be avoided in this formalism. The future Big Rip is avoided for a scale factor of the auxiliary metric as suggested from equation $42$ of \cite{lov9}. If there is a minimum length (and maximum density) at early times on homogeneous and isotropic space-times, then such predictions will lead to an alternative theory of the Big Bang \cite{lov10}.

A modified Friedman equation is obtained for the EiBi model as

\begin{equation}
\begin{split}
3H^2 = \frac{1}{k}\left[k \rho-1 + \frac{1}{3\sqrt{3}} \sqrt{(k \rho+1)(3-k \rho)^3} \right] \\ \times \left[ \frac{(k \rho+1)(3-k \rho)^3}{(3-k^2 \rho^2)^2}\right] 
\end{split}
\end{equation}
The minimum value for the scale factor is obtained as 
\begin{equation}
a_b \sim 10^{-32}(k)^\frac{1}{4} a_0 ,
\end{equation}

and the minimum length is predicted to be
\begin{equation}
a_b = \left(\frac{\rho_0}{\rho_b}\right)^{-4}
\end{equation}

Replacing  $\rho_b$ with $\rho_{crit}$ obtained from LQG, then at energy densities $\rho_b= 0.41 \rho_{pl}$, the universe will bounce back. Similarly, the minimum scale factor obtained from the EiBI calculations $a_b$ provides the same values as the minimal scale factor values predicted from the LQG. Both theories confirm the non-singular initial stages and singularity-free gravitational collapse. A tensor instability in the Eddington inspired Born-Infeld Theory of Gravity is reported in \cite{lov11}. The modified scale factor is obtained as

\begin{equation}\label{sam1}
a= a_b[1+tan^2{\Upsilon \eta}],
\end{equation}
where $\eta$ is  conformal time and 

\begin{equation}
\Upsilon = a_b\sqrt{\frac{2}{3|k|}}
\end{equation}

Relating the equations \ref{lqm1} and \ref{sam1} leads the scale factor as
\begin{equation}\label{sam2}
a = \left(\frac{3}{4}\rho_c t^2+1\right)^{\frac{1}{3}}[1+\tan^2(\left(\frac{3}{4}\rho_c t^2+1\right)^{\frac{1}{3}})\eta]
\end{equation}

Equation  \ref{sam2} is the modified scale factor obtained from loop quantum and EiBi solutions. The modified scale factor with the effect of EiBI solutions is obtained.

 \section{Discussion} 
 
 From the solutions of Eq. (\ref{lqm3.1}) is has been understood that the final stages of the universe will have the possibility to attain the critical energy density with values near to Planck density. Further increment of energy density is forbidden. Hence, the universe will bounce back to the formation of a new Aeon. This solution predicts the avoidance of a Big Rip. In this LQC-based work, modified solutions are implemented for the Wheeler-DeWitt solutions. Earlier, in Wheeler-DeWitt solution, the critical density parameter was not included.  From equations (\ref{m}) and (\ref{lqm1}) the scale factors for $k=0$ model have been equated. The solution for scale factor depends upon the selection of cosmological variables. From the WDW model standpoint, there is the possibility for the scale factor vanish as per the chosen values of the parameters. But LQC avoid such a scenario. Even at the singularity, LQC processes the non zero scale factor. Such results are the consequences of discreteness of quantized spacetime. Hence, the Hubble parameter is modified with loop quantum cosmology from equation (\ref{lqm2}).  Compared to the equation (\ref{k}),  the universe bounces back at singularity with matter bouncing energy densities, that are calculated from equation (\ref{lqm3}). The numerical predictions provided the value for the matter density at the bounce back, which is $\rho_{cric} \sim 0.41 \rho_{pl}$. The Wheeler-DeWitt quantum potential resolves time-varying scale factors. After the time-varying  parameter $N(t)$ is included in the quantum potential equation \ref{n} and it becomes equation \ref{lqm4}, obtained results of scale factors are compared with EiBI theory and classical supersymmetric cosmology. By comparing, it has been understood that the scale factor acts as a function of the trigonometric tangent function. EiBI-inspired modified scale factor is reported in equation \ref{sam2}. The minimum scale factor values are predicted from LQG and are consistent with the minimum scale factor, which is predicted from EiBI theory. The time-varying parameter $N(t)$ confirms the consistent value for the cosmological constant over the cosmological evolution. The cosmological constant which is proposed for the accelerated expansion of the universe behaves like a quantum potential. Hence, future bounce is available with increasing phantom field.  The FRW equations modified with the cosmological constant and renormalized quantized results are obtained from the LQC equations (\ref{lqa1}) and (\ref{lqa2})  respectively. The regular cosmological constant is renormalized within the LQC framework. This behaves as a function of the critical density.
 
 \begin{figure*}\label{img_2}
 \begin{center}
 \includegraphics[scale=0.5]{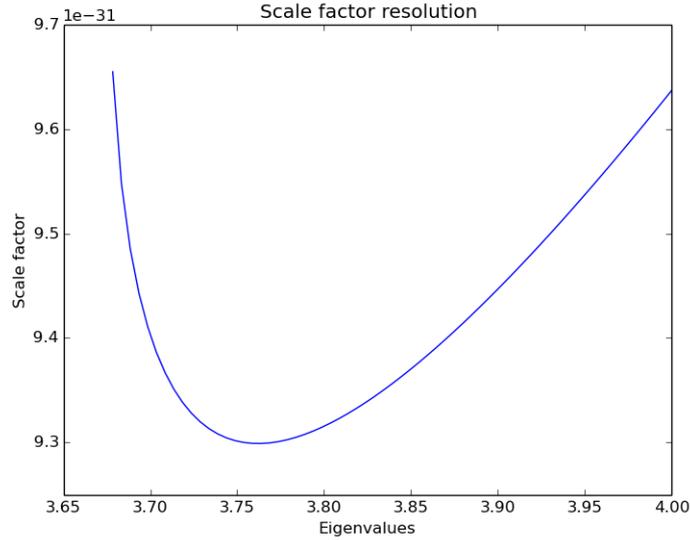}
 \caption{Non zero scale factor is predicted by LQG as a
 	function of Planck length. }
 \end{center}
 \end{figure*}

  \section{Conclusion}
 
  Although the WDW analysis attempted to quantize the singularity solutions, it also encountered a divergence problem. Loop quantum cosmological analysis confirms the existence of a non zero scale factor at the initial stages of the universe by transforming the scale factor as an operator. The author in \cite{phans1} discussed that the interaction between the phantom energy and dark matter lead to reduced density to a certain critical level. Consequently the Big Rip is avoided in the future universe. But we express an alternative to such conclusion; the universe will evolve even after the Big Rip, which means that the universe will continue its evolution in some other way. Interaction between the phantom energy and dark matter will be nonexistent while the coupling constant approaches zero. Then, the energy density of the phantom energy will continue to increase eternally. At later stages, the phantom energy density will be equal to the critical density. Subsequently, there is a possibility to bounce back for the future evolution. The role of $N(t)$ in equation (\ref{lqm4}) on cosmological evolution is to conserve the phantom energy. Hence the phantom energy remains unperturbed throughout the evolution. Also, the non-interacting solutions of phantom energy and dark matter hold the eternal nature of phantom energy (equation (\ref{noni1})). As the evolution continues even after the final singularity is approached, the viability of CCC is confirmed. The mapping between the scale factors of various models and various stages of the universe can be understood by the procedure of quantization. From the equation (\ref{ricci1}), if the values of cosmological parameters have specified values, such as $\rho =\rho_{crit}$ and $\omega<-1$, the universe will have negative curvature instead of zero curvature.  Hence, AdS kind of future universe might have appeared. Non zero values for the scale factor for the set of eigenvalues can be understood from the figure 2. The quantized scale factor never approaches zero at the initial stages of the universe, in spite of classical scale factor facing zero at an initial singularity. This could be the initial adjustment for the conformal mapping of initial singularity in conformal cyclic cosmology. Similarly, the Hamiltonian matter bounce is compared with the critical density parameter. The universe bounces back with the densities $\rho_{cri} \sim 0.41 \rho_{pl}$. The cosmological constant is quantized and renormalized within LQC formalism. Hence, inspired by the CCC model, the late time avoidance of Big Rip and continuing evolution can be held. Loop Quantum Cosmology provides some modifications on scale factor regularization, quantum potential, and Hamiltonian formulation. Additionally, included cosmological parameter on phantom energy, gives a consistent value for phantom energy throughout the evolution of the universe. Also, one can understand that the modified scale factor from EiBI theory can act as a function of the critical density.  An explanatory theory for the quantum emergence of the universe via conformal cyclic evolution is attempted and envisaged.

\bibliographystyle{spphys}     
\bibliography{ref2}

\begin{thebibliography}{10}
\providecommand{\url}[1]{{#1}}
\providecommand{\urlprefix}{URL }
\expandafter\ifx\csname urlstyle\endcsname\relax
  \providecommand{\doi}[1]{DOI \discretionary{}{}{}#1}\else
  \providecommand{\doi}{DOI \discretionary{}{}{}\begingroup
  \urlstyle{rm}\Url}\fi

\bibitem{cq1}
D.~An, K.A. Meissner, R.~Penrose, arXiv preprint arXiv:1808.01740  (2018)

\bibitem{vg1}
C.W. Misner, K.S. Thorne, J.A. Wheeler, D.I. Kaiser, \emph{Gravitation}
  (Princeton University Press, New Jersey,, 2017)

\bibitem{unb1}
D.~Tytler, J.M. O'Meara, N.~Suzuki, D.~Lubin, in \emph{Particle Phys. and the
  Universe} (World Scientific,, 2001), pp. 12--31

\bibitem{unm1}
J.~Khoury, B.A. Ovrut, P.J. Steinhardt, N.~Turok, Phys. Rev. D  (2001),
  \textbf{64}(12), 123522

\bibitem{unm3}
L.~Baum, P.H. Frampton, Phys. Rev. Lett.  (2007), \textbf{98}(7), 071301

\bibitem{lcd1}
J.~Wambsganss, P.~Bode, J.P. Ostriker, Astrophys. J. Lett.  (2004),
  \textbf{606}(2), L93

\bibitem{qv1}
E.P. Tryon, Nature  (1973), \textbf{246}(5433), 396

\bibitem{uvac1.5}
B.S. DeWitt, Phys. Rev.  (1967), \textbf{160}, 1113

\bibitem{uvac1}
D.~He, D.~Gao, Q.y. Cai, Phys. Rev. D  (2014), \textbf{89}(8), 083510

\bibitem{vc39}
A.~Vilenkin, Phys. Lett. B  (1982), \textbf{117}(1-2), 25

\bibitem{sin1.30}
V.~Taveras, Phys. Rev. D  (2008), \textbf{78}(6), 064072

\bibitem{sin1}
P.~Singh, Class. Quantum Grav.  (2009), \textbf{26}(12), 125005

\bibitem{sin1.18}
A.~Ashtekar, T.~Pawlowski, P.~Singh, Phys. Rev. D  (2006), \textbf{73}(12),
  124038

\bibitem{lc9}
I.~Agullo, A.~Corichi, in \emph{Springer Handbook of Spacetime}, ed. by
  A.~Ashtekar, V.~Petkov (Springer, Berlin,, 2014), pp. 809--839

\bibitem{lov5}
O.~Obreg{\'o}n, J.~Rosales, V.~Tkach, Phys. Rev. D  (1996), \textbf{53}(4),
  R1750

\bibitem{lov6}
L.P. Chimento, A.S. Jakubi, Int. J. Mod. Phys. D  (1996), \textbf{5}(01), 71

\bibitem{lov7}
L.A. Urena-Lopez, J. Cosmol. Astropart. Phys.  (2005), \textbf{2005}(09), 013

\bibitem{lov8}
C.~Escamilla-Rivera, O.~Obreg{\'o}n, L.A. Ure{\~n}a-L{\'o}pez, in \emph{AIP
  Conference Proceedings}, vol. 1256 (AIP, 2010), vol. 1256, pp. 262--266

\bibitem{phans1}
Z.K. Guo, Y.Z. Zhang, Phys. Rev. D  (2005), \textbf{71}(2), 023501

\bibitem{cq1.1}
R.~Penrose, in \emph{Proceedings of EPAC, Edinburgh, 2006}, ed. by C.~Prior
  (EPAC,Edinburgh,, 2006), p. p.2759

\bibitem{cq1.2}
R.~Penrose, \emph{Cycles of time: an extraordinary new view of the universe}
  (Random House,, 2010)

\bibitem{cq1.3}
R.~Penrose, Found. Phys.  (2018), \textbf{48}(10), 1177

\bibitem{lco1}
E.F. Bunn, N.~Sugiyama, arXiv preprint astro-ph/9407069  (1994)

\bibitem{cq7}
R.~Penrose, in \emph{AIP Conference Proceedings 11}, vol. 1446 (AIP, 2012),
  vol. 1446, pp. 233--243

\bibitem{efen1}
D.~Kramer, H.~Stephani, M.~MacCallum, E.~Herlt, Berlin  (1980)

\bibitem{phan01}
F.S. Lobo, Phys. Rev. D  (2005), \textbf{71}(8), 084011

\bibitem{phan02}
S.~Sushkov, Phys. Rev. D  (2005), \textbf{71}(4), 043520

\bibitem{27.deu2}
H.P. Nilles, Phys. Rep.  (1984), \textbf{110}(1-2), 1

\bibitem{28.deu2}
M.~Pollock, Phys. Lett. B  (1988), \textbf{215}(4), 635

\bibitem{29.deu2}
P.H. Frampton, Phys. Lett. B  (2003), \textbf{555}(3-4), 139

\bibitem{lov12}
L.~Fern{\'a}ndez-Jambrina, R.~Lazkoz, in \emph{Proceedings of the MG12 Meeting
  on General Relativity, Paris, 2009}, ed. by T.~Damour, R.~Jantzen, R.~Ruffini
  (World Scientific,Paris,, 2012), p. p.1887

\bibitem{luc1}
L.~Fernandez-Jambrina, R.~Lazkoz, Phys. Rev. D  (2004), \textbf{70}(12), 121503

\bibitem{luc2}
L.~Fernandez-Jambrina, R.~Lazkoz, Phys. Lett. B  (2009), \textbf{670}(4-5), 254

\bibitem{lov13}
C.~Escamilla-Rivera, Galaxies  (2016), \textbf{4}(3), 8

\bibitem{lov14}
L.P. Chimento, R.~Lazkoz, R.~Maartens, I.~Quiros, J. Cosmol. Astropart. Phys.
  (2006), \textbf{2006}(09), 004

\bibitem{lc9.93}
J.~Ganc, E.~Komatsu, Phys. Rev. D  (2012), \textbf{86}(2), 023518

\bibitem{sin1.39}
A.~Ashtekar, E.~Wilson-Ewing, Phys. Rev. D  (2008), \textbf{78}(6), 064047

\bibitem{lc8}
M.~Bojowald, Class. Quantum Grav.  (2009), \textbf{26}(7), 075020

\bibitem{lov1}
M.~Bojowald, Living Rev. Relativ.  (2008), \textbf{11}(1), 4

\bibitem{deu5.12}
C.~Cattoen, M.~Visser, Class. Quantum Grav.  (2005), \textbf{22}(23), 4913

\bibitem{deu5.13}
L.~Fernandez-Jambrina, R.~Lazkoz, Phys. Rev. D  (2006), \textbf{74}(6), 064030

\bibitem{cq2.14}
T.~Thiemann, Class. Quantum Grav.  (1998), \textbf{15}(5), 1281

\bibitem{cq2.10}
M.~Bojowald, Class. Quantum Grav.  (2000), \textbf{17}(6), 1509

\bibitem{uvac1.7}
N.~Pinto-Neto, Class. Quantum Grav.  (2013), \textbf{30}, 143001

\bibitem{uvac1.8}
N.~Pinto-Neto, F.~Falciano, R.~Pereira, E.S. Santini, Phys. Rev. D  (2012),
  \textbf{86}(6), 063504

\bibitem{uvac1.9}
S.P. Kim, arXiv preprint gr-qc/9703065  (1997)

\bibitem{uvac1.6}
A.~Vilenkin, Phys. Rev. D  (1994), \textbf{50}(4), 2581

\bibitem{uvac1.10}
S.W. Hawking, Nucl. Phys. B  (1984), \textbf{239}(1), 257

\bibitem{uvac1.14}
L.~Grishchuk, Class. Quantum Grav.  (1993), \textbf{10}(12), 2449

\bibitem{uvac1.15}
J.B. Hartle, S.~Hawking, T.~Hertog, J. Cosmol. Astropart. Phys.  (2014),
  \textbf{2014}(01), 015

\bibitem{uvac1.16}
D.~Coule, Class. Quantum Grav.  (2005), \textbf{22}(12), R125

\bibitem{lov2}
C.~Escamilla-Rivera, O.~Obreg{\'o}n, L.A. Ure{\~n}a-L{\'o}pez, J. Cosmol.
  Astropart. Phys.  (2010), \textbf{2010}(12), 011

\bibitem{phans3}
W.~Zimdahl, D.~Pavon, L.P. Chimento, Phys. Lett. B  (2001), \textbf{521}(3-4),
  133

\bibitem{lov3}
J.C. Fabris, S.V. Gon{\c{c}}alves, P.E. de~Souza, Gen. Relativ. Gravit.
  (2002), \textbf{34}(1), 53

\bibitem{lov4}
V.~Gorini, A.~Kamenshchik, U.~Moschella, Phys. Rev. D  (2003), \textbf{67}(6),
  063509

\bibitem{deu4}
P.~Singh, F.~Vidotto, Phys. Rev. D  (2011), \textbf{83}(6), 064027

\bibitem{phans2}
A.K. Yadav, AstroPhys. and Space Science  (2016), \textbf{361}(8), 276

\bibitem{lc9.26}
P.~Singh, Class. Quantum Grav.  (2012), \textbf{29}(24), 244002

\bibitem{vu1}
D.~Perlov, A.~Vilenkin, in \emph{Cosmology for the Curious} (Springer, 2017),
  pp. 333--341

\bibitem{deu2.12}
B.~Ratra, P.J. Peebles, Phys. Rev. D  (1988), \textbf{37}(12), 3406

\bibitem{deu2.27}
H.P. Nilles, Phys. Rep.  (1984), \textbf{110}(1-2), 1

\bibitem{deu2.28}
M.~Pollock, Phys. Lett. B  (1988), \textbf{215}(4), 635

\bibitem{lqsc1}
A.~Ashtekar, A.~Corichi, P.~Singh, Phys. Rev. D  (2008), \textbf{77}(2), 024046

\bibitem{lov9}
I.~Albarran, M.~Bouhmadi-L{\'o}pez, C.Y. Chen, P.~Chen, Phys. Lett. B  (2017),
  \textbf{772}, 814

\bibitem{lov10}
M.~Banados, P.G. Ferreira, Phys. Rev. Lett.  (2010), \textbf{105}(1), 011101

\bibitem{lov11}
C.~Escamilla-Rivera, M.~Banados, P.G. Ferreira, Phys. Rev. D  (2012),
  \textbf{85}(8), 087302

\end{thebibliography}

\end{document}